\documentclass[12pt,preprint]{aastex}

\newcommand{\neii}{[\ion{Ne}{2}]}
\newcommand{\neiii}{[\ion{Ne}{3}]}
\newcommand{\nev}{[\ion{Ne}{5}]}
\newcommand{\siii}{[\ion{S}{3}]}
\newcommand{\oi}{[\ion{O}{1}]}
\newcommand{\arii}{[\ion{Ar}{2}]}
\newcommand{\hii}{H$_{2}$}
\newcommand{\feii}{[\ion{Fe}{2}]}

\begin{document}

\title{A Large Mass of H$_{2}$ in the Brightest Cluster Galaxy in
Zwicky~3146}

\author{
E.~Egami\altaffilmark{1},
G.~H.~Rieke\altaffilmark{1},
D. Fadda\altaffilmark{2},
and
D.~C.~Hines\altaffilmark{3}
}

\altaffiltext{1}{Steward Observatory, University of Arizona, 933
  N. Cherry Avenue, Tucson, AZ~85721}

\altaffiltext{2}{NASA Herschel Science Center, Caltech, MC 100-22, 770
S.\ Wilson Avenue, Pasadena, CA 91125}

\altaffiltext{3}{Space Science Institute, 4750 Walnut Street, Suite
205, Boulder, CO~80301}

\begin{abstract}
 
We present the {\em Spitzer}/IRS mid-infrared spectrum of the
infrared-luminous ($L_{IR}=4 \times 10^{11} L_{\sun}$) brightest
cluster galaxy (BCG) in the X-ray-luminous cluster Z3146 ($z=0.29$).
The spectrum shows strong aromatic emission features, indicating that
the dominant source of the infrared luminosity is star formation.  The
most striking feature of the spectrum, however, is the exceptionally
strong molecular hydrogen (\hii) emission lines, which seem to be
shock-excited.  The line luminosities and inferred warm \hii\ gas mass
($\sim 10^{10} M_{\sun}$) are 6 times larger than those of NGC~6240,
the most \hii-luminous galaxy at $z \la 0.1$.  Together with the large
amount of cold \hii\ detected previously ($\sim 10^{11} M_{\sun}$),
this indicates that the Z3146 BCG contains disproportionately large
amounts of both warm and cold \hii\ gas for its infrared luminosity,
which may be related to the intracluster gas cooling process in the
cluster core.

\end{abstract}


\section{Introduction}

Zwicky~3146 (Z3146)\footnote{The NASA/IPAC Extragalactic Database
(NED) name of this cluster is ZwCl1021.0+0426.} is an X-ray-luminous
cluster of galaxies at a redshift of 0.29.  Its soft (0.1--2.4 keV)
X-ray luminosity \footnote{Throughout the paper, we adopt the
cosmological parameters of $\Omega_{M}=0.3$, $\Omega_{\Lambda}=0.7$,
and $H_{0}=70$ km s$^{-1}$.}  of $2 \times 10^{45}$ erg s$^{-1}$
\citep{Ebeling98,Boehringer00} is one of the largest known, and based
on the {\em ROSAT} data, Z3146 was thought to exhibit one of the most
massive cluster cooling flows \citep{Edge94}. Our analysis of {\em
Chandra} data confirms that the gas cooling rate is indeed large with
an estimated mass deposition rate of $\sim 300$ $M_{\sun}$ yr$^{-1}$
within a 50 kpc radius from the cluster center \citep{Egami06}.

The brightest cluster galaxy (BCG) is exceptionally active, showing a
blue continuum and luminous nebular emission lines \citep{Allen92,
Hicks05} and an infrared luminosity of $4 \times 10^{11}$ L$_{\sun}$
\citep{Egami06} .  The mid-infrared spectral energy distribution (SED)
suggests that the source of the large infrared luminosity is star
formation at a rate of $\sim$70 $M_{\sun}$ yr$^{-1}$, broadly
consistent with the mass deposition rate mentioned above. This may
indicate that the active star formation is caused by cooling cluster
gas accreting onto the BCG, i.e., the cooling flow \citep{Cowie77,
Fabian77}.  At the same time, the BCG seems to contain a radio AGN
that may contribute to its luminosity \citep{Egami06}.

To investigate the mid-infrared spectral properties in detail, we have
observed the BCG in Z3146 using the Infrared Spectrograph (IRS; Houck
et al. 2004) on board the {\em Spitzer} Space Telescope
\citep{Werner04}.

\section{Observations and Data Reduction}

The IRS spectra of the BCG in Z3146 covering 5--38 $\mu$m were taken
in the staring mode with the two low-resolution modules, Short-Low
(SL; $\Delta\lambda = 5.2 - 14.5 \mu$m) and Long-Low (LL;
$\Delta\lambda = 14.0 - 38.0 \mu$m), with a resolving power of $R \sim
64 - 128$.  The SL spectra were taken for 8 cycles of 60 s ramp
durations for both the 2nd (SL2: 5.2--8.7 $\mu$m) and 1st (SL1:
7.4--14.5 $\mu$m) orders while the LL spectra were taken for 8 cycles
of 120 s ramp durations for the 2nd order (LL2: 14.0--21.3$\mu$m) and
6 cycles for the 1st order (LL1: 19.5-- 38.0 $\mu$m).  The resultant
integration times are 960 s (SL1 and SL2), 1920 s (LL2), and 1440 s
(LL1), respectively, after coadding the spectra taken at the two nod
positions.  The slit width is 3\farcs7 for the SL and 10\farcs7 for
the LL module.

Starting with the reduced and coadded two dimensional spectral image
at each nod position, we performed sky subtraction in two passes,
first by taking the difference between the two nod positions (spectral
images targetting different orders were paired to avoid both the
positive and negative spectra falling in the same order) and second by
fitting and subtracting a linear fit to the background along each row
in the cross-dispersed direction (only fitting the pixels well away
from the source spectrum).  Then, pixels with an abnormal signal near
the source spectrum were manually masked out from the subsequent
processing.  From the background-subtracted and cleaned spectral
images, the source spectrum was extracted at each nod position using
the program
SPICE\footnote{\url{http://ssc.spitzer.caltech.edu/postbcd/spice.html}}
with extraction apertures reduced to 75\% of the default widths.  All
the spectra were combined by resampling them with a uniform resolving
power of 130 over the whole wavelength range.  Finally, the combined
spectrum was multiplied by 1.4 to match the IRAC 8.0 $\mu$m and MIPS
24 $\mu$m photometry.

\section{Results} 

\begin{figure}
\includegraphics[angle=90,scale=0.65]{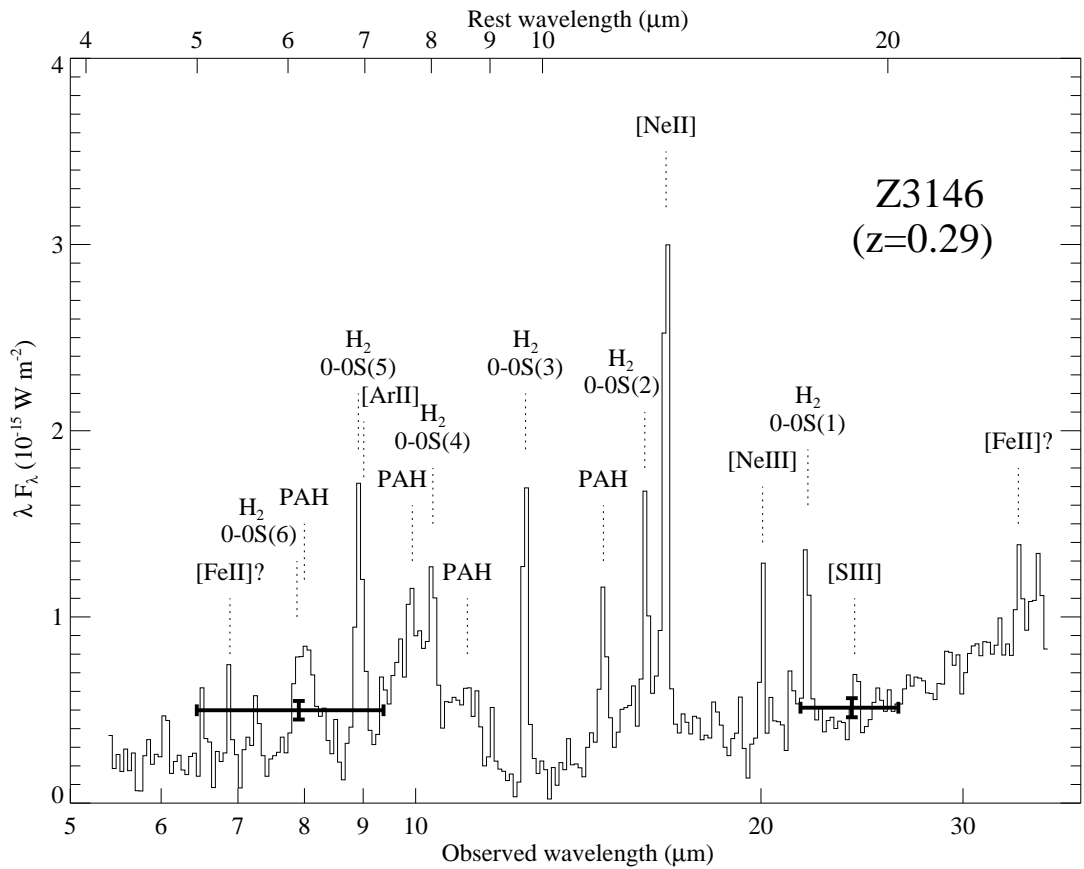}
\caption{Combined IRS spectrum of the BCG in Z3146.  The two points
with error bars show the IRAC 8.0$\mu$m and MIPS 24 $\mu$m photometry
from \citet{Egami06}.  The IRS spectrum has been scaled ($\times 1.4$)
to match these photometry points when the spectrum is integrated over
the corresponding passbands. \label{irs}}
\end{figure}

The final combined IRS spectrum of the BCG in Z3146 is shown in
Figure~\ref{irs}.  The measured line fluxes and rest-frame equivalent
widths are listed in Table~\ref{flux}.  Along the spatial direction,
the lines and continuum are unresolved (FWHM$\sim$2 pixels), which
sets an upper limit of 3\farcs6 in SL and 10\farcs2 in LL,
respectively, on the size of the mid-infrared emitting region.  This
is consistent with the fact that the Z3146 BCG is barely resolved at 8
$\mu$m with a spatial resolution of 2\arcsec\ \citep{Egami06}.

Strong aromatic (``PAH'') emission features are detected at 6.2, 7.7,
8.6, and 11.3 $\mu$m.  These features indicate that the mid-infrared
(and therefore presumably the far-infrared) luminosity of this galaxy
has a significant component of active star formation.

The spectrum also shows exceptionally strong emission lines from
molecular hydrogen (\hii).  We have securely detected the 0--0 S(1),
S(2), S(3), S(4), and S(5) pure rotational lines.  Although the 0--0
S(0) line is within the observed wavelength range ($\lambda_{obs} =
36.4 \mu$m), the noise in the LL1 band increases dramatically beyond
35 $\mu$m, preventing us from setting a meaningful upper limit on the
line flux.  The 0--0 S(6) line is blended with the bright 6.2 $\mu$m
PAH feature nearby.

A few atomic fine-structure emission lines are also detected in the
spectrum such as \neii\ 12.8 $\mu$m and \neiii\ 15.5 $\mu$m.  The
\siii\ 18.7 $\mu$m line is possibly detected as well, but because of
the low signal-to-noise ratio, we treat it as a non-detection here.
Also, no high-ionization line (e.g., \nev\ 15.6$\mu$m) is seen.  
The small \neiii/\neii\ line ratio of 0.2 is typical of a starburst
galaxy \citep{Thornley00} although we cannot rule out a significant
contribution to these neon lines from a shock-excited gas as
discussed below.  

There are possible detections of the \feii\ lines at 5.3 and 26.0
$\mu$m.  However, a longer integration time is needed to confirm the
reality of these lines.

\section{Discussion}

\subsection{PAH Features}

In terms of the strength of the 6.2 $\mu$m PAH feature, the Z3146 BCG
appears to be a typical starburst-dominated infrared luminous galaxy.
The rest-frame equivalent width of the feature (0.67 $\mu$m) as well
as its luminosity ratio to the total infrared luminosity ($5 \times
10^{-3}$) are both within the range found for starburst galaxies,
$\sim 0.6\pm0.1$ $\mu$m for the former \citep{Weedman05} and $(6.3 \pm
3.2) \times 10^{-3}$ for the latter \citep{Peeters04}, respectively.
Given the absence of any AGN-related features in the spectrum, the AGN
contribution to the infrared luminosity is likely to be small.

\subsection{Molecular Hydrogen Lines}


The most striking feature of the spectrum is the exceptionally strong
emission lines from molecular hydrogen.  The luminosities of the pure
rotational lines are on average 6.3 times larger than those of
NGC~6240, the most \hii-luminous galaxy at $z \la 0.1$.

\begin{figure}
\includegraphics[angle=90,scale=0.65]{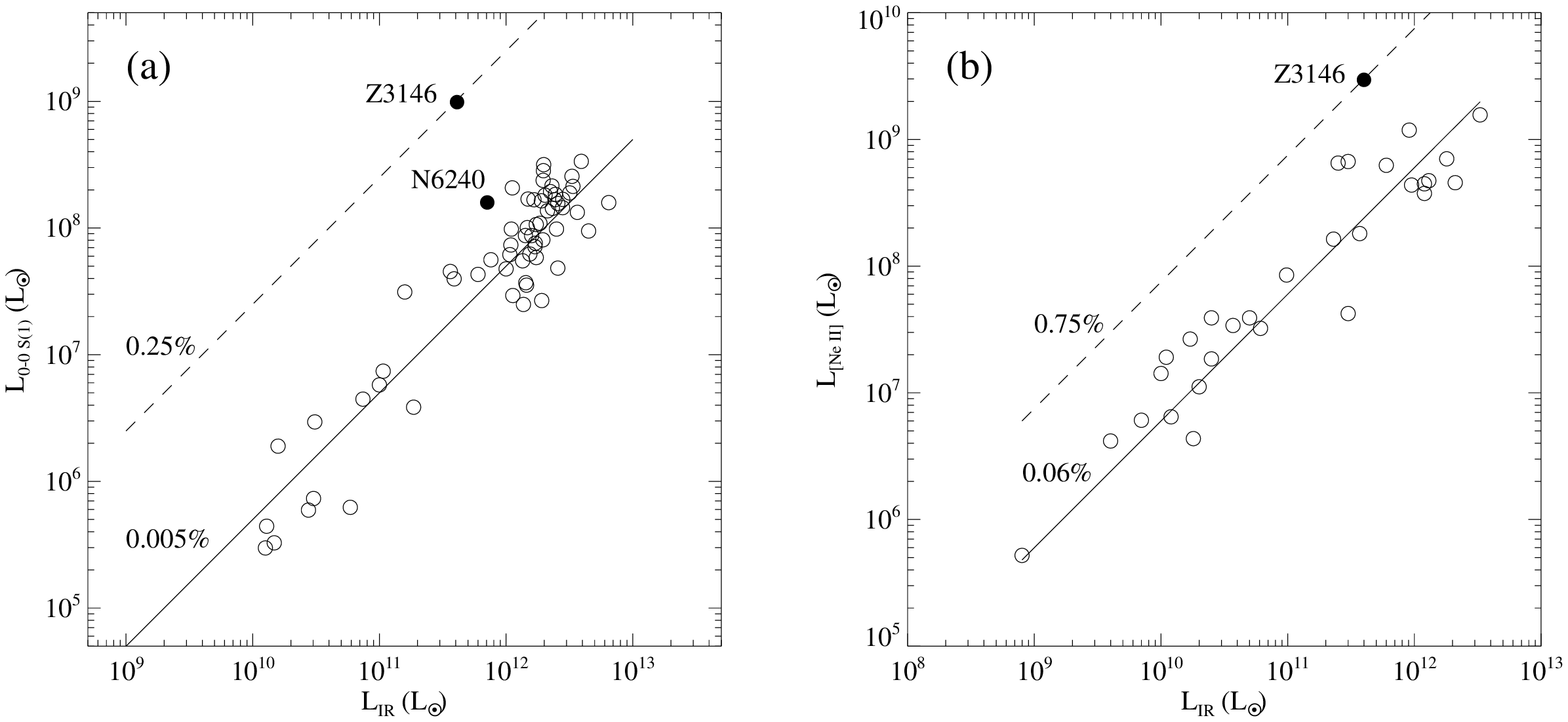}
\caption{(a) \hii\ 0--0 S(1) 17 $\mu$m line luminosity vs. infrared
luminosity ($L_{IR}$).  The solid and dashed lines correspond to
$L_{0-0S(1)} = $ 0.005 and 0.25\% of $L_{IR}$.  The \hii\ line
luminosities were taken from \citet{Armus06} (NGC~6240),
\citet{Rigopoulou02}, and \citet{Higdon06}.  Note that the \hii\
luminosity correlates with the infrared luminosity, so the sample
studied by \citet{Higdon06}, which extends beyond a redshift of 0.1
and therefore includes more infrared-luminous galaxies, contains
galaxies more \hii\ luminous than NGC~6240; (b) \neii\ 12.8 $\mu$m
line luminosity vs.\ $L_{IR}$. The solid and dashed lines correspond
to $L_{[Ne II]} = $ 0.06 and 0.75\% of $L_{IR}$.  The \neii\ line
luminosities were taken from \citet{Genzel98} while the infrared
luminosities were taken from \citet{Sanders03}. \label{lum_ratio}}
\end{figure}

The \hii\ line strengths are even more remarkable when compared with
the infrared luminosity of this BCG.  Figure~\ref{lum_ratio}a plots
the line luminosity of the \hii\ 0--0 S(1) emission line against the
infrared luminosity of various galaxies.  The \hii\ 0--0 S(1)/infrared
luminosity ratio of this BCG is extreme (0.25\%), almost an order of
magnitude larger than the ratio for NGC~6240 (0.03\%), and a factor of
50 above the trend line for the other galaxies (0.005\%).

\begin{figure}
\includegraphics[angle=90,scale=0.65]{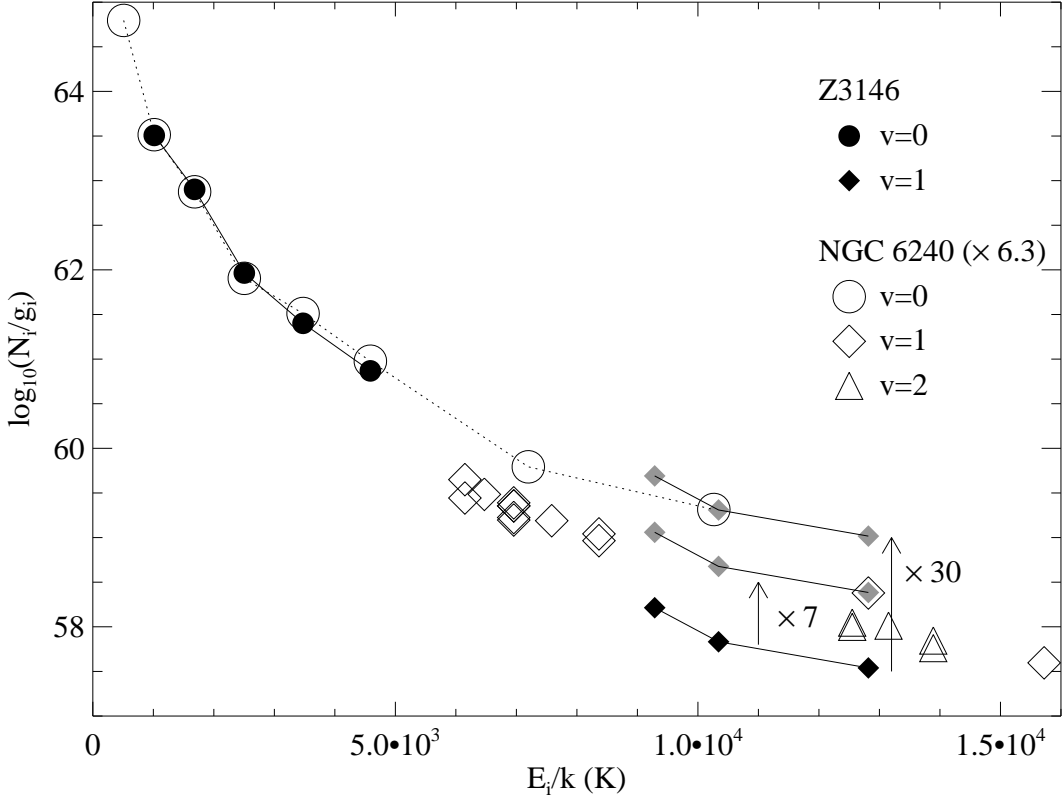}
\caption{\hii\ excitation diagram showing the level population of warm
\hii\ in the Z3146 BCG.  The level population in the ground
vibrational state ($v=0$) is based on our observation while the $v=1$
level population was derived from the data by \citet{Edge02}.  The X
axis is the \hii\ energy level expressed in temperature ($E_{i}$ is
the energy level of the $i$-th state; $k$ is the Boltzmann constant).
In the Y axis, $N_{i}$ refers to the {\em total} number of \hii\ in
the $i$-th energy state while $g_{i}$ is the statistical weight of
that state (ortho-para ratio was set to 3:1).  Assuming that the \hii\
lines are optically thin, $N_{i}$ was calculated as $N_{i} =
L_{i}/(A_{i} h \nu_{i})$, where $L_{i}$ is the observed line
luminosity, $A_{i}$ is the Einstein coefficient for that transition,
$h$ is the Planck constant, and $\nu_{i}$ is the frequency of the
line.  In comparison, the \hii\ level population of NGC~6240 is also
shown based on the published data
\citep{vdWerf96,Sugai97,Lutz03,Armus06}.  The data from
\citet{Sugai97} were plotted by assuming that the 1--0 S(1) line flux
agrees with that of \citet{vdWerf96}.  No other scaling was applied.
The solid line connects the $v=0$ and 1 states in the Z3146 BCG while
the dotted line connects the $v=0$ states in NGC~6240. \label{h2}}
\end{figure}

Figure~\ref{h2} shows the \hii\ level population in the Z3146 BCG.
For comparison, the level population in NGC~6240 is overplotted after
being scaled by a factor of 6.3.  The level populations are quite
similar in shape between the two galaxies in the lowest energy states
($v=0$; $J=3-7$).  These lowest energy states are almost always
thermalized (i.e., collisionally excited), and therefore the
similarity suggests that although the total mass of warm \hii\ differs
between the two galaxies, the fractional \hii\ mass distribution as a
function of temperature is similar up to a gas temperature of
$\sim$960 K (i.e., the \hii\ excitation temperature derived for the
$v=0$, $J=6$ and 7 states using the 0--0 S(4) and S(5) lines).

Assuming that this similarity of the level population extends down to
the $v=0, J=2$ state, from which the 0--0 S(0) line originates, we
estimate the total warm \hii\ mass in the Z3146 BCG to be $\sim
10^{10} M_{\sun}$, i.e., 6.3 times larger than that of NGC~6240 ($1.6
\times 10^{9} M_{\sun}$ by \citet{Armus06}).  This mass is dominated
by the $\sim$160 K component, a temperature derived from the 0--0
S(0)/0--0 S(1) line ratio measured for NGC~6240.  The cold \hii\ mass
derived from the CO observation is also large, $\sim 10^{11} M_{\sun}$
\citep{Edge03}, but the fraction of warm \hii\ mass is high,
$\sim$10\%, similar to the 15\% fraction found in NGC~6240
\citep{Armus06}.

Figure~\ref{h2} also shows that the \hii\ level population of the
Z3146 BCG in the $v=1$ vibrational state is highly suppressed (by a
factor of 30) with respect to that in the $v=0$ state when the
NGC~6240 $v=0$ level population is used as a template.  Although part
of this suppression is due to the slit loss with the near-infrared
spectroscopy as well as internal dust extinction, these effects, even
when combined, are unlikely to be sufficient.  For example, the size
of the mid-infrared emitting region is 2\arcsec\ at most as already
mentioned, so the 1\farcs5 slit used to measure the $v=1-0$
near-infrared \hii\ lines \citep{Edge02} should capture most of the
total line fluxes.  Also, in this BCG, the extinction derived from the
Balmer decrement is $A_{V}=0.6 $ mag \citep{Crawford99}, and when
corrected for this amount, the H$\alpha$-derived star formation rate
agrees well with the infrared-derived star formation rate
\citep{Egami06}, suggesting that the extinction cannot be much larger.
In fact, near-/mid-infrared \hii\ emission lines are known to show
little extinction in infrared-luminous galaxies
\citep{Goldader97,Higdon06}.

This suppression of the $v=1$ state suggests that the $v=1$ level
population in the Z3146 BCG is subthermal as is the case with NGC~6240
\citep{Egami98,Lutz03}.  Although the cause of such a subthermal level
population is not clear, one possibility is that the \hii\ molecules
are excited by a continuous (i.e., C-type) shock, which can produce
subthermal level populations in the low vibrational states
\citep[e.g.,][]{Kaufman96,LeBourlot02}.  Note that alternative
excitation mechanisms such as a jump (J-type) shock or far-UV
radiation would produce a $v=1$ level population at or above the
thermal level with respect to the $v=0$ state
\citep[e.g.,][]{Burton92}.  Figure~\ref{h2} shows that only a factor
of $\sim$7 scaling is required to bring up the $v=1$ level population
in the Z3146 BCG to that of NGC~6240.  This correction factor is much
easier to account for than the factor of 30 mentioned above, which
would be required if the $v=1$ state is thermalized.  Shock excitation
is also consistent with the previous near-infrared spectroscopic
studies of nearby cooling-flow cluster BCGs
\citep[e.g.,][]{Jaffe01,Wilman02}.

\subsection{Atomic Fine-Structure Lines}

The \neii\ 12.8 $\mu$m line of this BCG is also strong.
Figure~\ref{lum_ratio}b plots the \neii\ 12.8 $\mu$m luminosity
against the infrared luminosity for a sample of nearby infrared-bright
galaxies.  As shown by \citet{Sturm02}, there is a good correlation
between the two luminosities among the starburst and Seyfert galaxies,
and the Z3146 BCG again stands out by having an exceptionally strong
\neii\ 12.8 $\mu$m line for its infrared luminosity (0.75\%),

This overluminous \neii\ emission line may also arise from a shock.
However, the C-type shock invoked above to explain the \hii\ level
population cannot produce a substantial amount of the \neii\ line
because the shocked gas in a C-type shock is largely neutral and
molecular.  To produce the \neii\ line, the shock must be a fast ($\ga
50$ km s$^{-1}$) J-type shock, that can dissociate and ionize the
shocked gas \citep[e.g.,][]{Hollenbach89}.  The Z3146 BCG has an
H$\alpha$ velocity width of $\sim600$ km s$^{-1}$ \citep{Allen92}, so
such a fast shock is plausible.  

This probably means that various types of shocks may coexist in the
Z3146 BCG.  A similar conclusion was reached for NGC~6240 by
\citet{Lutz03} to account for the strong \oi\ 63 $\mu$m line, which is
also a sign of a J-type shock.  Indeed, such a coexistence of the C-
and J-type shocks is also seen in some Galactic outflow sources, in
which both strong \hii\ and \neii\ lines are detected
\citep[e.g.,][]{Lefloch03}.  Shock models can produce a wide range of
\neiii/\neii\ line ratios depending on various shock conditions, and
therefore it is possible to produce the starburst like \neiii/\neii\
ratio with a strong shock \citep[e.g.,][]{Binette85, Hartigan87}.
\feii\ lines may also be used to constrain the shock conditions if
their strengths are measured accurately.

With respect to the \neii\ 12.8 $\mu$m line, the \siii\ 18.7 $\mu$m
line is abnormally weak (\siii/\neii\ $<$0.05).  Such a low
\siii/\neii\ ratio is often seen with ultraluminous infrared galaxies
(ULIRGs), indicating that the gas phase abundance of sulfur is
significantly below solar \citep{Genzel98}.  Starburst galaxies also
show this sulphur ``deficiency'' although the \siii/\neii\ ratios are
less extreme, and this may suggest that the sulphur is depleted onto
dust grains in high-metallicity star-forming systems \citep{Verma03}.

\section{Conclusions}

We present the {\em Spitzer}/IRS mid-infrared spectrum of the
infrared-luminous BCG in the X-ray luminous cluster Z3146.  The strong
aromatic features indicate that the dominant source of the infrared
luminosity is star formation.  The most striking feature of the
spectrum, however, is the exceptionally strong \hii\ emission lines.
To our knowledge, this BCG has the most luminous \hii\ pure-rotational
emission lines and most massive warm \hii\ gas component ($\sim
10^{10} M_{\sun}$) known.  Together with the large amount of cold
\hii\ detected previously ($\sim 10^{11} M_{\sun}$), this indicates
that the Z3146 BCG contains disproportionately large amounts of both
warm and cold \hii\ gas for its infrared luminosity.  Since the parent
cluster Z3146 has an exceptionally large gas cooling rate in the core,
these massive warm and cold \hii\ components may be related to the
intracluster gas cooling process.

\acknowledgments

We would like to thank Dr.\ G.~Neugebauer and the anonymous referee
for their comments, which improved the paper.  This work is based on
observations made with the {\em Spitzer} Space Telescope, which is
operated by the Jet Propulsion Laboratory, California Institute of
Technology under a contract with NASA (contract number \#1407).
Support for this work was provided by NASA through an award issued by
JPL/Caltech (contract number \#1255094).

\clearpage

\begin{deluxetable}{lccc}
\tablecaption{Z3146 BCG line fluxes \label{flux}}
\tablewidth{0pt}
\tablehead{
\colhead{Line/Feature} & \colhead{$\lambda_{\rm rest}$} & \colhead{Flux}  &
\colhead{EW$_{rest}$} \\
\colhead{}       & \colhead{($\mu$m)}  & \colhead{(10$^{-17}$ W m$^{-2}$)} &
\colhead{($\mu$m)}
}
\startdata                                                              
PAH                     &   6.2   &   3.0$\pm$0.2  &  0.67$\pm$0.05 \\  
H$_{2}$ \phn $0-0$ S(5) &   6.9   &   2.1$\pm$0.1  &  0.49$\pm$0.02 \\  %
\arii\tablenotemark{a}  &   6.99  &   $\sim$0.6    &  $\sim$0.15    \\  %
PAH                     &   7.7   &   6.9$\pm$0.2  &  1.72$\pm$0.06 \\  
H$_{2}$ \phn $0-0$ S(4) &   8.0   &   0.8$\pm$0.1  &  0.19$\pm$0.02 \\  
PAH                     &   8.6   &   1.2$\pm$0.2  &  0.31$\pm$0.04 \\  
H$_{2}$ \phn $0-0$ S(3) &   9.7   &   2.3$\pm$0.1  &  1.72$\pm$0.05 \\  
PAH                     &  11.3   &   1.3$\pm$0.1  &  0.47$\pm$0.07 \\  
H$_{2}$ \phn $0-0$ S(2) &  12.3   &   1.2$\pm$0.1  &  0.26$\pm$0.02 \\  
\neii                   &  12.8   &   4.2$\pm$0.1  &  1.34$\pm$0.03 \\  
\neiii                  &  15.6   &   0.8$\pm$0.1  &  0.30$\pm$0.04 \\  
H$_{2}$ \phn $0-0$ S(1) &  17.0   &   1.4$\pm$0.1  &  0.55$\pm$0.04 \\  
\siii                   &  18.7   &   $<0.4$       &  $< 0.2$       \\  
\enddata

\tablenotetext{a}{The \arii\ 6.99 $\mu$m line was not detected
directly, but its presence was set to the maximum level consistent
with the S(5) line shape.}

\end{deluxetable}

\clearpage

\clearpage

\clearpage

\end{document}